\documentclass[journal]{IEEEtran}

\usepackage{amssymb,amsmath}
\usepackage{paralist}
\usepackage{graphicx}

\usepackage{color}
\usepackage{arydshln}
\definecolor{rltblue}{rgb}{0,0,0.75}
\usepackage{cite,manfnt}

\newcommand{\F}{\mathbf{F}}

\newcommand{\C}{\mathcal{C}}
\newcommand{\N}{\mathcal{N}}
\newcommand{\G}{\mathcal{G}}

\newtheorem{theorem}{\textbf{Theorem}}
\newtheorem{lemma}[theorem]{\textbf{Lemma}}
\newtheorem{definition}[theorem]{\textbf{Definition}}

\newtheorem{remark}[theorem]{\textbf{Remark}}

\newtheorem{example}[theorem]{\textbf{Example}}

\newcommand{\nix}[1]{}

\begin{document}
\title{Network Protection Codes: Providing Self-healing in Autonomic Networks Using Network Coding
\thanks{This paper was presented in part at the IEEE Globecom 2008
Conference, New Orleans, LA, December 1-4, 2008~\cite{aly08i}.\nix{ This research was supported in part by grants CNS-0626741 and
CNS-0721453 from the National Science Foundation, and a gift from Cisco Systems.}}
}
\author{
\authorblockN{Salah A. Aly ~~~~ and ~~~~ Ahmed E. Kamal\\}
\authorblockA{Department of Electrical and Computer Engineering \\Iowa State University,  Ames, IA 50011, USA\\Emails:\{salah,kamal\}@iastate.edu}\\
\begin{small}\end{small} }
%
\markboth{Submitted to IEEE Journal of Selected Areas in Communications ({J-SAC}),  2008.} {Aly, Kamal, \MakeLowercase{\textit{ Submitted to   \MakeUppercase{IEEE}  Journal of Selected Areas in Communciations (J-SAC),  2008.}}} \maketitle

\begin{abstract}
Agile recovery from link failures in autonomic communication networks is
essential to increase robustness, accessibility, and reliability of data
transmission.
However, this must be done with the least amount of protection
resources, while using simple management plane functionality.
Recently, network coding has been proposed as a solution
to provide agile and cost efficient network self-healing
against link  failures, in a manner that does not require data rerouting,
packet retransmission, or failure localization, hence leading to
simple control and management planes. To achieve this, separate paths have to be
provisioned to carry encoded packets, hence requiring either the
addition of extra links, or reserving some of the resources for this
purpose.

In this paper we introduce autonomic self-healing strategies for
autonomic networks in order to protect against link failures.
The strategies are based on network coding and reduced capacity, which
is a technique that we call \emph{network protection codes} (NPC).
In these strategies, an autonomic network is able to provide
self-healing from various network failures
affecting network operation. The techniques improve service and
enhance reliability of autonomic communication.

\emph{Network protection codes} are  extended to provide self-healing from multiple
link failures in autonomic networks.
Although this leads to reducing the network capacity, the network capacity
reduction is asymptotically small in most cases of practical
interest. We provide implementation aspects of the proposed strategies. We present bounds and \emph{network protection code} constructions.
Furthermore tables of the best known self-healing codes are presented. Finally, we study the construction of such codes over the binary field.
The paper also develops an Integer Linear Program formulation to
evaluate the cost of provisioning connections using the proposed
strategies, and uses results from this formulation to show that it is
more resource efficient from 1+1 protection.
\end{abstract}

\section{Introduction}\label{sec:intro}

Today's communication networks are becoming complex to the degree that
the management of such networks has become a major task of network
operation.
Therefore, the use of network autonomy such that the management
functionality and its complexity, is moved to within the network has
become the preferred approach, hence giving rise to what is known as
autonomic networks~\cite{schmid06}.
Autonomic networks are self-managed, and they
are efficient, resilient, evolvable, through self-protection,
self-organizations, self-configurations, self-healing
and self-optimizations (see for example~\cite{gu08,IBM01,tesauro07}  and
the references therein). Therefore an autonomic network promotes the
autonomy of operational networks with minimum human involvements.
However, it is also important not to overload the management plane of
autonomic networks to the degree that the management functionality
consumes significant amount of computing and communication resources.
This paper addresses the self-functionality  in autonomic networks, and
introduces a technique to provide self-healing that results in simplifying
the management plane, as well as the control plane.
The technique uses reduced capacities and network coding.

Network coding is  a powerful tool that has been used to increase
the throughput, capacity, and performance of communication
networks~\cite{soljanin07,yeung06}. It offers benefits in terms of
energy efficiency, additional security, and reduced delay. Network
coding allows the intermediate nodes not only to forward packets
using network scheduling algorithms, but also encode/decode them
using algebraic primitive operations
(see~\cite{ahlswede00,fragouli06,soljanin07,yeung06} and references
therein).

One application of network coding that has been proposed recently is
to provide network protection against link failures in overlay
networks~\cite{kamal06a,kamal08b}. This is achieved by transmitting
combinations of data units from multiple connections on a backup
path in a manner that enables each receiver node to recover a copy
of the data transmitted on the working path in case the working path
fails. This can result in recovery from failures without data
rerouting, hence achieving agile protection. Moreover, the sharing
of network protection resources between multiple connections through the
transmission of linear combinations of data units results in
efficient use of protection resources. This, however, requires the
establishment of extra paths over which the combined data units are
transmitted. Such paths may require the addition of links to the
network under the Separate Capacity Provisioning strategy (SCP), or
that paths be provisioned using existing links if using the Joint
Capacity Provisioning strategy (JCP), hence reducing the network
traffic carrying capacity.

Certain networks can allow extra transmissions and the addition of
bandwidth, but they do not allow the addition of new transmission lines.  In this scenario,
one needs to design efficient data recovery schemes.    In this paper,
we propose such an approach in which we use network coding to
provide agile, and resource efficient protection against link
failures, and without adding extra paths. The approach is based on
combining data units from a number of sources, and then transmitting
the encoded data units using a small fraction of the bandwidth
allocated to the connections, hence disposing of the requirement of
having extra paths. In this scenario, once a path fails, the
receiver can recover the lost packets easily from the neighbors by
initiating simple queries.

\goodbreak
Previous solutions in network survivability approaches using network
coding focused on providing backup paths to recover the data
affected by the
failures~\cite{kamal06a,kamal07a,kamal07b}. Such
approaches include 1+N, and M+N protections. In 1+N protection, an
extra secondary path is used to carry combinations of data units
from N different connections, and is therefore used to protect N
primary paths from any single link failure. The M+N is an extension
of 1+N protection where M extra secondary paths are needed to
protect multiple link failures.

In this paper, we introduce autonomic self-healing and healing-protection network strategies based on network coding and reduced capacity.
In these strategies, an autonomic network is able to provide
self-healing from various network failures.
The techniques improve services and
enhance reliability of autonomic communication.   We define the concept of
\emph{network protection codes} similar to error-correcting codes that are widely
used in channel coding~\cite{huffman03,macwilliams77}.  Such codes aim to provide better
provisioning and data recovery mechanisms~\cite{aly08i}.

\bigbreak

The new contributions in this paper are stated as follows:
\begin{compactenum}[i)]
\item We introduce a self-healing strategy using
network coding and a reduced capacity strategy instead of
using dedicated paths.

\item
We provide a new scheme to protect against a single link failure  in
autonomic networks. The scheme  is extended to protect against multiple
link failures.

\item
We develop a theoretical foundation of \emph{protection codes}, in which
the receivers are able to recover data sent over $t$ failed links
out of $n$ primary links.

\item The developed protection strategies are achieved over the binary
field, hence the encoding and decoding operations are done using XOR
operation.

\end{compactenum}

\medskip

This paper is organized as follows. In Section~\ref{sec:relatedwork}
we briefly state the related work and previous solutions to the
network protection problem against link failures.  In Section~\ref{sec:networkmodel} we
present the network model and problem definition.
Sections~\ref{sec:singlefailure} and~\ref{sec:multiplefailures}
discuss single and multiple link failures  and how to protect these
link failures using reduced capacity and network coding. In
Section~\ref{sec:analysis} we give analysis of the general case of
$t \ll n$ link failures. Sections~\ref{sec:codeconstruction} and~\ref{sec:bestcodes} present code constructions  and bounds on the \emph{network protection code} parameters. In Section~\ref{sec:ilp} we present an integer
linear program to find the optimal provisioning under the proposed
scheme.
Section \ref{sec:cost} introduces some numerical results based on the
ILP and a comparison between 1+1 protection and the proposed scheme.
The paper is concluded in
Section~\ref{sec:conclusion}.

\noindent \textbf{Notations:} We  fix the notation throughout the paper.
Let $n$, $k$,  $m$, and $t$ be the number of total connections, working
paths, protection paths, and failures, respectively, where $n=k+m$ and
$t \leq k$.
Let $L_i$ be a connection from a sender $s_i$ to a receiver $r_i$. Let
$c_i$ be the unit capacity of the connection $L_i$ if it carries plain
data (data without coding). $\F_2$ is a finite field with two elements
$\{0,1\}$.
An $[n,k,d_{min}]_2$ is a \emph{network protection code} defined
over $\F_2$ that has $n$ connections, $k$ working paths, $n-k=m$
protection paths, and recovers from $t=d_{min}-1$ failures, where $d_{min}$ is the minimum distance of the code.

\section{Related Work}\label{sec:relatedwork}

In this section we will state the related work in network protection strategies against link failures, and linear codes that are used for erasure channels. We define the concept of
\emph{network protection codes} similar to error-correcting codes that are widely
used in erasure channel coding~\cite{huffman03,macwilliams77}.

\subsection{Revolution Networks Using Network Coding}

Network coding is  a powerful tool that has been used to increase
the throughput, capacity, and performance of communication
networks~\cite{soljanin07,yeung06}. Network coding assumes that the network nodes not only can forward incoming messages/packets, but also can encode, decode them. It offers benefits in terms of
energy efficiency, additional security, and reduced delay~
(see~\cite{ahlswede00,fragouli06,soljanin07,yeung06} and references
therein).
Practical aspects of network coding have been investigated
in~\cite{chou03}, and
bounds on the network coding capacity are investigated in~\cite{aly07e,ramamoorthy05}.

\subsection{Protection against Failures Using Network Coding}
In~\cite{kamal06a}, the author introduced a 1+N protection model in
optical mesh networks using network coding over p-cycles. The author
suggested a model for protecting $N$ connections  from a set of
sources to a set of receivers in a network with $n$ connections,
where only one connection might fail.
Hence, the suggested model can protect
against a single link failure in any arbitrary path connecting a
source and destination.
In~\cite{kamal07a}, the author extended the previous model to
protect against multiple link failures. It is shown that protecting against
$m$ failures, at least $m$ p-cycles are needed.  The idea was to
derive $m$ linearly independent equations to recover the data sent
from $m$ sources.
In~\cite{kamal07b}, the author extended the protection model
in~\cite{kamal06a} and provided a GMPLS-based implementation of a
link protection strategy that is a hybrid of 1+N and 1:N. It is
claimed that the hybrid 1+N link protection provides protection at
higher layers and with a speed that is comparable to the speed
achieved by the physical layer implementations. In addition, it has
less cost and much flexibility.

In this paper, we provide a new technique for protecting a network
against failures using \emph{protection codes} and \emph{reduced
capacity}, and for the network to recover from such failures in
an agile manner.
The benefits of our approach are that:
\begin{compactenum}[i)]
\item
It allows receivers to recover the lost data without data
rerouting, data retransmission or failure localization, hence
simplifying the control and management planes.
\item
It has reasonable computational complexity and does not require adding
extra paths {or reserving backup paths}.
\item
{At any point in time,}
all $n$ connection paths have full capacity except at one path in case of
protecting against a single link failure and $m < n$ paths in case
of protecting against $ t\leq m $ link failures.
\item The working and protection paths capacities are distributed among each other for fairness.
\end{compactenum}

\section{Network Model}\label{sec:networkmodel}

Let $\mathcal{G}=(V,E)$ be a graph which represents the network topology.  $V$ is a set of network nodes and $E$ is
a set of edges. Let there be $n$ unidirectional connections,
and let $S \subset V$ be the set of sources $\{s_1,...,s_n
\}$ and $R \subset V\backslash S$ be the set of receiver nodes $\{
r_1,...,r_{n} \}$ of the $n$ connections in $ \mathcal{G}$. The case of
$S \cap R \neq \phi$ can be easily incorporated in our model. Two nodes $u$  and $v$ in $ V$  are connected by an edge $(u,v)$ in $E$ if there is a
direct connection between them.  We assume that the sources are
independent of each other, meaning they can only send messages and
there is no correlation between them. For simplicity, we will assume
that a path exists between $s_i$
and $r_i$, {and it is disjoint from the path between $s_j$ and $r_j$, for $j \neq i$}.

The network model $\N$ can be described in the following assumptions.

\begin{compactenum}[i)]
\item  Let
$\N$ be a network with a set of sources $S=\{s_1,s_2,\ldots,s_n\}$
and a set of receivers $R=\{r_1,r_2,\ldots,r_n\}$, where $S \cup
R \subset V$.

\item Let $L$ be a set of links  $L_1,L_2,\ldots,L_n$ such that there
is a link $L_i$ if and only if there is a connection path between
the sender $s_i$ and receiver $r_i$, i.e., $L_i$ corresponds to the path
\begin{eqnarray}
    \{(s_i,w_{1i}),(w_{1i},w_{2i}),\ldots,(w_{(\lambda)i},r_i) \},\end{eqnarray} where
    $1\leq i\leq n$ and $(w_{(j-1)i},w_{ji}) \in E$, for some integer $\lambda \geq 1$.
Hence
we have $|S|=|R|=|L|=n$. The n connection paths are pairwise link
disjoint.

\item Every source $s_\ell$ sends a packet with its own $ID_{s_\ell}$ and data
 $x_\ell$ to the receiver $r_\ell$, so
 \begin{eqnarray}
packet_{s_\ell}=(ID_{s_\ell},x_\ell, \delta),
 \end{eqnarray}
where $\delta$ is the round number of the
source packet $packet_{s_\ell}$.

 \item All packets belonging to the same round
are sent in the  same round slot. The senders will exchange the rule of
sending plain and encoded data for fairness, as will be illustrated
below.

\item All links carry uni-directional data from sources to
receivers.
\item  We consider the scenario where the cost
of adding a new path is higher than just combining messages in an
existing path, or there is not enough resources to provision dedicated
paths in the network.
\end{compactenum}

We can define the unit capacity $c_i$ of a link $L_i$ as follows.

\begin{definition}\label{def:capacitylink}
The unit capacity of a connecting path $L_i$ between $s_i$ and $r_i$ is defined
by \begin{eqnarray} c_i=\left\{
      \begin{array}{ll}
        1, & \hbox{$L_i$ is an \emph{active working path};} \\
        0, & \hbox{otherwise.}
      \end{array}
    \right.
\end{eqnarray}
What we mean by an \emph{active}  path is that the receiver is
able to receive and process unencoded signals/packets throughout this path. Hence, the protection path is assumed to be inactive.
The total capacity of $\N$ is given by the summation of all active path
capacities, divided by the number of paths.
\end{definition}

This means that each source $s_i$  can send a maximum of one packet per unit time
on a link $L_i$. Assume that all links have the same capacity.
One can also always assume that a
source with a large rate can be divided into a set of sources, each
of which has a unit link capacity.
\goodbreak

\begin{figure}[t]
 \begin{center} 
  \includegraphics[height=5cm,width=8.3cm]{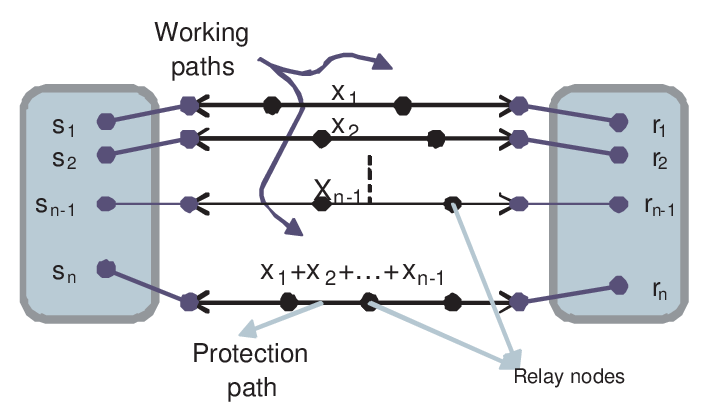}
  \caption{Network protection against a single path failure using reduced capacity and network coding.
   One path out of  $n$ primary paths  carries encoded data. The black points represent various other relay nodes}\label{fig:npaths}
\end{center}
\end{figure}

The following definition describes the \emph{working} and
\emph{protection} paths between two network switches as shown in Fig.~(\ref{fig:npaths}).

\begin{definition}
The\emph{ working paths} in a network with $n$ connection paths carry
traffic under normal operations. The data on these paths are sent
without encoding. The \emph{Protection paths} in our proposed scheme
carry encoded data from other sources. A
protection scheme ensures that data sent from the sources will reach the
receivers in case of failure on the working paths.
\end{definition}

Our goal is to provide an agile and resource efficient self-healing method for $n$
connections without adding extra paths.
Unencoded data is sent
over a path $L_i$ without adding extra paths.
but by possibly reducing the source rates slightly.
Linear combinations of data units are sent on these paths alternately,
and by using the reduction in working path capacities.
The linear combinations are used to recover from failures.

 	\smallbreak

Clearly, if all paths are active then the total capacity of all
connections is $n$.

In general, the total normalized capacity of the network for the active and failed paths is
computed by
\begin{eqnarray}
C_\N=\frac{1}{n}\sum_{i=1}^n c_i.
\end{eqnarray}

\bigbreak

\section{Protecting Networks Against  A Single Link Failure}\label{sec:singlefailure}
In this section we study the problem of protecting a set of
connections against a single link failure in a network $\N$ with a
set of sources $S$ and a set of receivers $R$. This problem has been
studied in~\cite{kamal06a,kamal07a} by provisioning a path that is
link disjoint from all connection paths, and passes through all
sources and destinations.  All source packets are encoded in one
single packet and transmitted over this path. The encoding is
dynamic in the sense that packets are added and removed at each
source and destination.

Assume  that
\typeout{the  assumptions about our network model $\N$, and the
abstraction graph $\G$ presented in Section~\ref{sec:networkmodel}
hold. We know that if there is an active link $L_i$ between $s_i$
and $r_i$, then the capacity $c_i$ is the unit capacity.
Let us
consider the case where}
every source $s_i$ has its own message $x_i$.
Also, source $s_j$ forms the encoded
data $y_j$
which is defined by
\begin{eqnarray}
y_j=x_1\oplus\ldots \oplus x_{i \neq
j} \oplus\ldots \oplus x_n
\label{eqn:yj}
\end{eqnarray}
where the sum is over the finite field
$\F_2=\{0,1\}$.
In this case,
the symbol $\oplus$ is the XOR operation.

Source $s_i$, for $i \neq j$,
sends a packet to the receivers $r_i$, which is given by

 \begin{eqnarray}
packet_{s_i}=(ID_{s_i},x_i,
\delta).
 \end{eqnarray}

On the other hand,

source $s_j$ sends a packet that will carry the encoded data $y_j$ to the receiver $r_j$  over the link $L_j$,
 \begin{eqnarray}
packet_{s_j}=(ID_{s_j},y_j, \delta).
 \end{eqnarray}

Now we consider the case where
there is a single failure on link $L_k$.
Therefore, we have two cases:
\begin{compactenum}[i)]
\item
If $k=j$,  the link $L_j$  has a failure, and
the receiver $r_j$ does not need to query any other node
since link $L_j$ carries encoded data that is only
used for protection.
All other receiver nodes receive their data correctly on links which
have not failed.

\item
If $k \neq j$, then the receiver $r_k$ needs to query the other $(n-1)$
nodes in order to recover the lost data $x_k$ over the failed link
$L_k$.
The reason is that $x_k$ exists either at  $r_j$, and
it requires information of all other receivers.
$x_k$ can be recovered by adding all other $n-1$ data units.
The recovery is implemented by adding $y_j$ and all $x_i$ for $i\neq j$,
and $i \neq k$.
This follows from Equation~(\ref{eqn:yj}).

\end{compactenum}

This shows that only one single receiver needs to perform $(n-2)$ addition operations in order to recover its data if its link fails. In
other words, all other receivers will receive the transmitted data
from the senders of their own connections with a constant operation
$O(1)$.

The following example illustrates the plain and encoded data transmitted from five senders to five receivers.

\medskip

\begin{example}
Let $S$ and $R$ be two sets of senders and receivers, respectively, in the network model $\N$. The following scheme explains the plain and encoded data sent in five consecutive rounds from the five senders to the five receivers.
\begin{eqnarray}
\begin{array}{|c|ccccc|c|c|}
\hline
cycle& &~~~~1&&&&2&3\\
\hline
rounds&1&2&3&4&5&\ldots&\ldots\\
\hline
\hline
s_1 \rightarrow r_1 &y_1&x_1^1  &x_1^2 &x_1^3 &x_1^4& \ldots&\ldots\\
s_2 \rightarrow r_2 &x_2^1&y_2  &x_2^2 &x_2^3&x_2^4&\ldots&\ldots\\
s_3 \rightarrow r_3 &x_3^1  &x_3^2&y_3 &x_3^3&x_3^4&\ldots&\ldots \\
s_4 \rightarrow r_4     &x_4^1 &x_4^2 & x_4^3&y_4&x_4^4&\ldots&\ldots\\
s_5 \rightarrow r_5     &x_5^1 &x_5^2 & x_5^3&x_5^4&y_4&\ldots&\ldots\\
\hline
\end{array}
\end{eqnarray}
The encoded data $y_j$, for $1 \leq j \leq 5$,  is sent as

\begin{eqnarray}
y_j=\sum_{i=1}^{j-1} x_i^{j-1}+\sum_{i=j+1}^5 x_i^{j}.
\end{eqnarray}
We notice that every message has its own round. Hence the protection data is distributed among all paths for fairness.
\end{example}

\subsection{Network Protection Codes (NPC) for a Single Link Failure}
We can define the set of sources that will send encoded packets by
using constraint matrices.  We assume that there is a \emph{network
protection code} $\C \subseteq \F_2^{n}$ defined by the constraint
matrix

\begin{eqnarray}\label{eq:G}
G\!\!=\!\! \left[ \begin{array}{ccccccccc}1&0&\ldots&0&1\\ 0&1& \ldots&0&1\\
\vdots&\vdots&\vdots&\vdots&\vdots\\
 0&0&\ldots&1&1\end{array}\right]_{(n-1)\times n}\!\!
\end{eqnarray}

\medskip

\begin{figure}[t]
 \begin{center} 
  \includegraphics[scale=0.7]{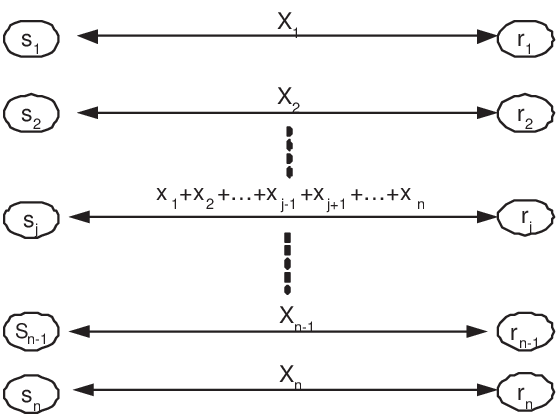}
  \caption{Network protection against a single link failure using reduced capacity and network coding. One connection out of  $n$ primary working paths  carries encoded data, i.e. protection path. There are $n-1$ active working paths carry plain data.}\label{fig:nnodes}
\end{center}
\end{figure}

Without loss of generality, in Matrix~(\ref{eq:G}), for $1 \leq j \leq n-1$, the  column vector $(\begin{array}{ccccc}
g_{1j}&g_{2j}&\ldots&g_{(n-1)j}\end{array})^T$ in $\F_2^{n-1}$ corresponds
to (n-1) sources, say for example the sources $s_1,s_2,\ldots,s_{n-1}$, that will send (update) their
values to (n-1) receivers, say i.e., $r_1,r_2,\ldots,r_{n-1}$. Also,
there exists one source that will send encoded data,
e. g., source $n$ in the above matrix.
The row vector
$(\begin{array}{ccccc} g_{i1}&g_{i2}&\ldots&g_{in}\end{array})$ in
$\F_2^n$ determines  the channels $L_1,L_2,\ldots,L_n$.

The weight of a row in $G$ is the number of nonzero elements.  We define $d_{min}$ to be the minimum weight of a row in $G$. Put differently

\begin{eqnarray}
d_{min}=\min \{ |g_{ij} \neq 0, 1\leq j \leq n|, 1\leq i \leq n-1\}
\end{eqnarray}
Hence, since every row in $G$ has weight of two, $d_{min}=2$.
\smallskip

We can now define the \emph{network protection code} that will protect a single path failure as follows:

\goodbreak

\begin{definition}
An $[n,n-1,2]$  \emph{network protection code} $\C$ is a ${n-1}$-dimensional
subspace of the space $\F_2^{n}$ defined by the generator systematic matrix
$G$ and is able to recover from a single network failure of an arbitrary
path  $L_i$.
\end{definition}

This means that an $[n,n-1,2]$ code over $\F_2$ is a code that encodes $(n-1)$ symbols into $n$ symbols and detects (recovers from) a single path failure.
We note that the \emph{network protection codes} NPC are also error correcting codes that can be used for erasure channels. The positions of errors (failures) are known.
\begin{remark}The number of failures that can be recovered by an NPC is equal to the minimum distance of the code minus one, i.e., $t=d_{min}-1$. Sometimes we refer to NPC by  the number of failures $t$, otherwise they are defined by the minimum distance $d_{min}$ as shown in Table~(\ref{table:bchtable3}).
\end{remark}

\smallskip

In general we will assume that the code $\C$ defined by the
generator matrix $G$ is known for every source $s_i$ and every
receiver $r_i$. This means that every receiver will be able to
recover the data $x_i$ if the link $L_i$ fails, provided that $L_i$ is
active in the sense defined in Definition~(\ref{def:capacitylink}).
Hence, the rows of the generator matrix $G$ are the basis for the code $\C$. We assume that the positions of the failures are known. Furthermore,
every source node has a copy of the code $\C$. Without loss of generality, the protection systematic matrix
among all sources is given by:
\medskip

\begin{eqnarray}\label{eq:matrixXi}
\begin{array}{c||cccccc}
&L_1&L_2&\cdots&L_{n-1} &L_n&\\ \hline \hline
s_1&x_1&0&\cdots&0&x_{1}\\
s_2&0&x_2&\cdots&0&x_{2}\\
\vdots&\vdots&\vdots&\cdots&\vdots&\vdots\\
s_{n-1}&0&0&\cdots&x_{n-1}&x_{n-1}\\
\hline \hline total&x_1&x_2&\ldots&x_{n-1}&y_n\\
\end{array}
\end{eqnarray}
\medskip
where $y_n$ is the protection value collected from every source $s_i$
that will be encoded at source $s_n$, for
all $1\leq i \leq n-1$.  Put differently, we have
\begin{eqnarray}\label{eq:protectionvalues}
y_n=\sum_{i=1}^{n-1}  x_{i}
\end{eqnarray}
where
the summation operation is defined by the XOR operation.

\medskip

In a general scenario,
the system operates in cycles, where each cycle consists of $n$ rounds,
such that at round $1\leq j \leq n$ of a cycle we have
\begin{eqnarray}\label{eq:protectionvalues2}
y_{j}=\sum_{i=1, i \neq j}^n x_{i}
\end{eqnarray}
where the round number of a packet $x_i$ is not shown for simplicity
with the understanding that it is the first packet in the source's
output queue.
We assume that every source $s_i$ has a buffer that   stores its
value $x_i$ and can also send the protection value $y_{j}$. Hence in the channel $L_j$, $s_j$ prepares
a  packet $packet_{s_j}$ that contains the value
\begin{eqnarray}
packet_{s_j}= (ID_{s_j}, y_{j}, \delta),
\end{eqnarray}
and sender $s_i$ for $i \neq j$  will send its data  $x_i$ in a  $packet_{s_i}$ over the channel $L_i$ defined as follows
\begin{eqnarray}
packet_{s_i}= (ID_{s_i}, x_{i}, \delta),
\end{eqnarray}
In general each source will send $(n-1)$ packets containing plain data,
and exactly one packet contain encoded data in all $n$ rounds. The
transmission will be repeated in cycles, hence every cycle has $n$ rounds.

 \medbreak

Recovery from a single path failure is summarized by the next two
lemmas.

\medskip

\begin{lemma}
Encoding the data from sources $S\backslash \{s_j\}$ at a source
$s_j$ in the network $\N$ is  enough to protect against a single
path failure.
\end{lemma}

\medskip

\begin{lemma}
The total number of encoding operations needed to recover from a
single link failure in a network $\N$ with $n$ sources is given by
 {$(n-2)$} and the total number of transmissions  is $n$.
\end{lemma}

The previous lemma guarantees the recovery from  a single arbitrary
link failure.

\medbreak

\begin{lemma}\label{lem:capacitysinglelink}
In  the network model $\N$, through out each cycle,  the average network capacity of
protecting against a single link failure using reduced capacity and
network coding is given by $(n-1)/n$.
\end{lemma}
\begin{proof}
\begin{inparaenum}[i)] \item We know that every source $s_i$ that sends the data
$x_i$ over a working path $L_i$ has capacity $c_{i}=1$. \item Also, the source $s_j$
  sends  the encoded data $y_{j}$ at  different slots,
has an inactive capacity.
\item The source $s_j$ is not fixed among all nodes $S$,
however, it is  rotated periodically over all sources for
fairness. On average one source of the $n$ nodes will reduce its
capacity. This shows the capacity of $\N$ as stated.
\end{inparaenum}
\end{proof}

\section{Protecting Networks Against Multiple
Link Failures}\label{sec:multiplefailures}

In the previous section we introduced a strategy for self-healing
from single link failure for autonomic
networks.

However, it was shown
in~\cite{markopoulou04} through an experimental study that about
$\%30$ of the failures of the Sprint backbone network are multiple
link failures. Hence, one needs to design a general strategy against
multiple link failures for  the purpose of self-healing.

In this section we will generalize the above strategy
to protect against $t$ path
failures using \emph{network protection codes} (NPC) and the reduced
capacity. We have the following assumptions about the channel model:
\begin{compactenum}[i)]
\item We assume that any $t$ arbitrary paths may fail and they
may or may not be correlated.
\item Locations of the failures are known, but they are arbitrary among $n$ connections.
    \item

In order to protect $n$ working paths, $k$ connection must carry plain data, and $m=n-k$ connections must carry encoded data.
\item We do not add extra
protection paths, and every source  node is able to encode  the incoming
packets independently.
\item We consider the encoding and decoding operations are performed over $\F_2$.

\end{compactenum}
In Sections~\ref{sec:bestcodes} and~\ref{sec:codeconstruction}  we will show the connection between error correcting
codes that are used for erasure channels and the proposed  \emph{network protection codes}~\cite{huffman03,macwilliams77}.

Assume that the notations in the previous sections hold. Let us
assume a network model $\N$ with $t>1$ path failures.  One can
define a\emph{ protection code} $\C$ which protects $n$ links as shown in
the systematic matrix $G$ in~(\ref{eq:Gmultiple}). In general, the  systematic matrix $G$
defines the source nodes that will send encoded  messages and source
nodes that will send only plain message without encoding. In order to protect $n$ working paths, $k$ connection must carry plain data, and $m=n-k$ connections must carry encoded data.

\goodbreak

The
generator matrix of the NPC for multiple link failures is given by:

\medskip

\begin{eqnarray}\label{eq:Gmultiple}
G= \left[
\begin{array}{cc}
\begin{array}{cccc}1&0&\ldots&0\\
0& 1&\ldots&0\\
\vdots&\vdots&\vdots&\\
 0&0&\ldots&1\\
 \end{array}&\begin{array}{cccc}|&p_{11}&\ldots&p_{1m}\\
|&p_{21}&\ldots&p_{2m}\\
|&\vdots&\vdots&\vdots\\
|&p_{k1}&\ldots&p_{km}\\
 \end{array} \\\\
 \underbrace{\hskip 0.7in}^{\mbox{identity matrix $k\times k$}} &\underbrace{\hspace{0.7in}}^{\mbox{ Submatrix } P_{k \times m}}\\
\end{array}
  \right],
\end{eqnarray}
where $p_{ij} \in \F_2$
\medskip

The matrix $G$ can be rewritten as
\begin{eqnarray}
G= \Big[\mbox{ } I_k \mbox{ } \mid  \mbox{ } \textbf{P} \mbox{ }
\Big],
\end{eqnarray}
where $\textbf{P}$ is the sub-matrix that defines the redundant data
$\sum_{i=1}^k p_{ij}$ to be sent to a set of sources for the purpose
of self-healing from
link failures. Based on the above matrix, every source $s_i$ sends
its own message $x_i$ to the receiver $r_i$ via the link $L_i$. In
addition $m$ links out of the $n$ links will carry encoded data.
Let $d_{min}$ be the minimum distance (minimum weight) of a nonzero vector in the matrix $G$.

\begin{definition}\label{defn:mfailuresCode}
An $[n,k,d_{min}]_2$ \emph{network protection code} $\C$ is a $k$-dimensional subspace of
the space $\F_2^{n}$ that is able to  recover from all network failures
up to $t=d_{min}-1$.
\end{definition}

\smallskip

In general the \emph{network protection code} (NPC), which protects against
multiple path failures, can be defined by a generator matrix $G$
known for every sender and receiver.
Also, there exists a parity check matrix $H$ corresponds to $G$ such that $GH^T=0$.
We will restrict ourselves in this work
for NPC that are generated by a given generator  matrix $G$ in the
systematic. In addition, we will assume that the \emph{protection codes } are
defined by systematic matrices defined over
$\F_2$~\cite{huffman03,macwilliams77}. An $[n,k,t]_2$ NPC code is also
an $[n,k,d_{min}]_2$, where  $t=d_{min}-1$.

Without loss of generality, at one particular round and cycle, the protection matrix (scheme) among all sources is given by

\medskip

\begin{eqnarray}
\begin{array}{c||cccccccc}
&L_1&L_2&\cdots&L_k &L_{k+1}&L_{k+2}&\ldots&L_{n}\\
 \hline \hline
s_1\!\!&x_1&0&\cdots&0&p_{11} x_{1}&p_{12} x_{1}&\!\!\ldots&\!\!p_{1m} x_{1}\\
s_2\!\!&0&x_2&\cdots&0&p_{21} x_{2}&p_{22} x_{2}&\!\!\ldots&\!\!p_{2m} x_{2}\\
\vdots\!\!&\vdots&\vdots&\cdots&\vdots&\vdots&\!\!\vdots&\!\!\vdots\\
s_{k}\!\!&0&0&\cdots&x_k&p_{k1} x_{k}&p_{k2} x_{k}&\!\!\ldots&\!\!p_{km} x_{k}\\
\hline \hline T\!\!&x_1&x_2&\ldots&x_k&y_1& y_2&\!\!\ldots&\!\!y_m\\
\end{array}
\end{eqnarray}
\medskip

We ensure that $k=n-m$ paths $L_1,L_2,\ldots,L_k$ have full capacity and they carry the
plain data $x_1,x_2,\ldots,x_k$. Also, all other $m$ paths have inactive capacity, in which
they carry the encoded data $y_1,y_2,\ldots,y_m$. In addition, the $m$
links are not fixed, and they are chosen alternatively between the
$n$ connections.

\subsection{Encoding and Recovery Operations}
We shall illustrate how the encoding and recovery operations are achieved at the sources and receivers, respectively.

\goodbreak

 \noindent \textbf{Encoding Process.} The network encoding
process at the set of senders are performed in a similar manner as
in Section~\ref{sec:singlefailure}. Every source $s_i$ has a copy of
the  systematic matrix $G$ and it will prepare a packet along with its ID in
two different cases. First, if the source $s_i$ will send only its
own data $x_i$ with a full link capacity, then

\begin{eqnarray}
packet_{s_i} = (ID_{s_i}, x_{i}, \delta).
\end{eqnarray}

Second,
if $\mathbb{S}$ is the set of sources sending encoded messages, then

\begin{eqnarray}
packet_{s_j} = (ID_{s_j},  \sum_{\ell=1, s_\ell
\notin \mathbb{S}}
^k p_{\ell j}x_\ell, \delta),
\end{eqnarray}
where $p_{\ell j} \in \F_{2}$.

The transmissions are sent in rounds. Therefore, the senders will
alternate the role of sending plain and encoded data for fairness.

\noindent \textbf{Recovery Process.} The recovery process is done as
follows. Assume $t$ failures occur, then a system of linearly
independent equations  of $t$ variables
(corresponding to the data lost due to the failed paths)
can be solved. The  $packet_{s_i}$ arrives at a receiver $r_i$
with an associated round number, $\delta$.
The receiver $r_i$ at time slot
$n$ will detect the signal in the link $L_i$. If the link $L_i$
fails, then $r_i$ will send a query to other receivers in $R
\backslash \{r_i\}$ asking for their received data.
Assume there are
$t $ path failures. Then we have three
cases:
\begin{compactenum}
\item
All $t$ link failures have occurred in links that  carry encoded
packets, i.e., $packet_{s_j} = (ID_{s_j},
\sum_{\ell=1, s_\ell\notin \mathbb{S}}^k p_{lj} x_\ell, \delta)$. In this case no recovery operations are needed.
\item All $t$ link failures have occurred in links that do not carry encoded
packets, i.e., $packet_{s_i} = (ID_{s_i}, x_{i}, \delta)$. In
this case, one receiver that carries encoded packets, e.g., $r_j$, can send
$n-m-1$ queries to the other receivers with active links asking for
their received data. After this
process, the receiver $r_j$ is able to decode all messages and will
send individual messages to all receivers with link failures to pass
their correct data.

\item
All $t$ link failures have occurred in arbitrary links. This case is
a combination of the previous two cases and the recovery process is
done in a similar way. Only the lost data on the working paths need to be recovered.
\end{compactenum}

\medskip

 The proposed
network protection scheme using distributed capacity and coding is able to recover up
to  $t \leq d_{min}-1 $ link failures (as defined in Definition~(\ref{defn:mfailuresCode}))
among $n$ paths and it has the following advantages:
\begin{compactenum}[i)]
\item
$k=n-m$ links have full capacity and their sender nodes have the same
transmission rate.
\item
The $m$ links that carry encoded data are dynamic (distributed) among all $n$
links. So, no single link $L_i$ will always suffer from reduced
capacity.

\item
The encoding process is simple once every sender  $s_i$ knows the
NPC.
\item The recovery from link failures is done in a dynamic and simple
way. Only one receiver node needs to perform the decoding process
and it passes the data to other receivers that have link failures.
\end{compactenum}

\section{Capacity Analysis}\label{sec:analysis}

We shall provide theoretical analysis regarding our \emph{network
protection codes}. One can easily compute the number of paths needed
to carry encoded messages to protect against $t$ link failures, and
compute the average network capacity. The main idea behind NPC is to
simplify the encoding operations at the sources and the recovery
operations at the receivers. The following lemma demonstrates the
average normalized capacity of the proposed network model $\N$ where $r$ failures occur.

\begin{lemma}\label{lem:capacitymulitplelinks}
Let $\C$ be a \emph{network protection code} with parameters $ [n,n-m,d_{min}]$ over $\F_2$. Let $n$ and $m$ be the number of sources (receivers) and number of
connections carrying encoded packets, respectively, the average normalized  capacity of the
network $\N$ is given by
\begin{eqnarray}
(n- m )/n.
\end{eqnarray}
\end{lemma}
\begin{proof}
At one particular round, we have $m$ protection paths that carry encoded data. Hence there are $n-m$ working paths that carry plain data. The result is a direct consequence by applying the normalized capacity definition.
\end{proof}

\medskip

\begin{remark}
In the  network protection model $\N$, in order to protect $t$
network disjoint link failures, the minimum distance $d_{min}$ of the protection code must be at least $t+1$.
\end{remark}

The previous remark ensures that the maximum number of failures that can be recovered is $d_{min}-1$, where $d_{min}$ is the minimum distance of the \emph{network protection code}.  For simplicity, we denote a NPC defined over $\F_2$ by $[n,n-m,d_{min}]_2$ unless stated otherwise.

For example one can use the Hamming codes with parameters $[2^r-1,2^r-r-1,3]_2$ to recover from two failures.
One can also puncture or extend these codes to reach the required length, i.e., number of connection, see~\cite{huffman03} for deriving new codes from known codes by puncturing, extending, shortening those codes.  $[7,4,3]_2$, $[15,11,3]_2$, and $[63,57,3]_2$ are examples of Hamming codes that protect against two link failures.  The protection code $[15,11,3]$ has $15$ connections among them are $11$ working paths and $4$ protection paths, in addition the minimum distance is $3$ and the code protects $2$ link failures.

Another example is the BCH codes with arbitrary design distance, i.e., $[n,k, d_{min}\geq \delta]_2$.
It is well known that
the minimum distance of a BCH code is greater than or equal to its designed distance.
References $[15,11,3]_2$, $[31,26,3]_2$ and $[63,56,3]_2$ are examples
of BCH codes that protect up to two link failures. Also,  $[15,8,5]_2$, $[31,21,5]_2$ and $[48,36,5]_2$ are examples of BCH codes against four link failures~\cite{huffman03,macwilliams77}.
In the next section we will include tables of the best known \emph{network  protection codes}.

\section{Code Constructions and Bounds}\label{sec:codeconstruction}

Assume we have $n$ established connections in the network model $\N$.
The goal is to design a good \emph{protection code}  that protects $t$ failures. What we
mean by a good \emph{protection code} is that for given number of connections $n$ and failures $t$, it has large number of working paths. Hence the protection code has a high performance.
In addition, we establish bounds on the
\emph{network protection code} parameters in the next section.

One way to achieve our goal is to design codes with arbitrary minimum distances. The reader can consult any introductory coding theory book, for example~\cite{macwilliams77,huffman03}. In this case a BCH code with designed distance $d$  and length $n$ can be used to deploy this goal.

We shall quickly review the essential construction of nonprimitive narrow-sense BCH codes that will be used in the next section. Let $q$ be a prime power, and  $n,\mu$ and $d$ be positive integers such that $\gcd(q,n)=1$, and $2 \leq d \leq n$.
Furthermore, $\mu$ is the multiplicative order of $q$ modulo $n$. Let $\alpha$ be
a primitive element in $\F_{q^\mu}$. A nonprimitive narrow-sense BCH
code $\mathcal{C}$ of designed distance $d$ and length $q^{\lfloor \mu/2\rfloor} <n \leq q^\mu-1$  over
$\F_{q}$ is a cyclic code with a generator monic polynomial $g(x)$
that has $\alpha, \alpha^2, \ldots, \alpha^{d-1}$ as zeros,
\begin{eqnarray}
g(x)=\prod_{i=1}^{d -1} (x-\alpha^i).
\end{eqnarray}
Thus,  $c$ is a codeword in $\mathcal{C}$ if and only if
$c(\alpha)=c(\alpha^2)=\ldots=c(\alpha^{d-1})=0$. The parity check matrix
of this code can be defined as
\begin{eqnarray}\label{bch:parity}
 H_{bch} =\left[ \begin{array}{ccccc}
1 &\alpha &\alpha^2 &\cdots &\alpha^{n-1}\\
1 &\alpha^2 &\alpha^4 &\cdots &\alpha^{2(n-1)}\\
\vdots& \vdots &\vdots &\ddots &\vdots\\
1 &\alpha^{d-1} &\alpha^{2(d-1)} &\cdots &\alpha^{(d-1)(n-1)}
\end{array}\right].
\end{eqnarray}

If the minimum distance of this code is $d_{min} \geq d$, then  the code can recover up to $d_{min}-1$ failures. In this case the number of connections that will carry plain data is given by:
\begin{equation}\label{eq:npdimension}
k\leq n-\mu\lceil (d-1)(1-1/q)\rceil.
\end{equation}

But this is an upper bound in the dimension of the NPC, aka, the number of working connections that carry plain data. Therefore, we seek a result to determine the exact dimension. Fortunately, this can be obtained when the designed distance of BCH codes are bounded.
The following Theorem enables one to determine the dimension in closed form for
BCH code of small designed distance.

\goodbreak

\begin{theorem}\label{th:bchnpdimension}
Let $q$ be a prime power and $\gcd(n,q)=1$, with $q^\mu \equiv 1 \mod n$.
Then a
narrow-sense BCH code of length $q^{\lfloor \mu/2\rfloor} <n \leq q^\mu-1$ over
$\F_q$ with designed distance $d$ in the range $2 \leq d \le
d_{\max}= \min\{ \lfloor nq^{\lceil \mu/2 \rceil}/(q^\mu-1)\rfloor,n\}$, has
dimension of
\begin{equation}\label{eq:npdimension}
k=n-\mu\lceil (d-1)(1-1/q)\rceil.
\end{equation}
\end{theorem}
\begin{proof}
See~\cite[Theorem 10]{aly07a}.
\end{proof}

For small designed distance $d$ we can exactly compute the minimum distance
of the BCH code, see Tables~(\ref{table:singledoublefailures}), (\ref{table:bchtable2}),  and~(\ref{table:bchtable3}). Consequently, determine the dimension of the protection code. This
helps us to compute the number of failures that the \emph{network protection
code} can recover. In practical cases, the number of failures $t$ is
small in comparison to the number of connections $n$ that makes it easy
to exactly compute the parameters of the \emph{network protection codes}.
Theorem~(\ref{th:bchnpdimension}) made it explicit straightforward to
derive the exact parameters of  NPC based on BCH codes.

We shall give many families of NPC codes derived from BCH codes over $\F_2$.
One final thing is that one can also start by a code for a  given length
$n$, and will be able to puncture, shorten, or extend this code,
see\cite[Chapter 1.]{huffman03}.
This will dramatically change the
number of working and protection paths and failures which the code can
recover.
\begin{table}[t]
\caption{Best known   \emph{network protection codes} against single and double link failures}
\label{table:singledoublefailures}
\begin{center}
\begin{tabular}{|l|l|l|l|}
\hline    n& m &code&type  \\
 \hline
 &&&\\
7&3&$[7,4,3]_2$&Hamming code\\
10&4&$[10,6,3]_2$&Linear code\\
15&4&$[15, 11, 3]_2$& Hamming code\\
19&7&$[19,12,3]_2$&Extension construction\\
23&8&$[23,15,3]_2$& Extension construction \\
25&5&$[25,20,3]_2$&Linear code\\
31&5&$[31, 26, 3]_2$& Hamming code\\
39& 8&$[39, 31,3]_2$&Extension construction \\
47&9&$[ 47,38 ,3]_2$&Extension construction\\
63&6&$[63, 57,3]_2$&Hamming code\\
71& 8&$[71,63,3]_2$&Matrix construction\\
79&9&$[79,70,3]_2$&Extension construction\\
95&10&$[95,85,3]_2$&Extension construction\\
127&7&$[127,120,3]_2$&Hamming code\\
\hline
\end{tabular}
\end{center}
\end{table}
\subsection{Bounds on the Code Parameters}
Bounds on the code parameters are needed to measure its performance and
error recovery and detection capabilities. For a given code parameters
length $n$ and dimension $k$, we establish a bound on the minimum
distance of the \emph{protection codes} derived in the previous section.

The most well-known upper bounds on error-correcting codes over
symmetric and erasure channels are the Singleton and Hamming
bounds~\cite{macwilliams77,huffman03}. The Singleton bound establishes
the relationship between the length, dimension, and minimum distance of
the code parameters, i.e. n, k, and $d_{min}$. However, it does not
specify the connection between code parameters and the alphabets size q.
The packing bound, known as Hamming bound, takes in consideration the
codes parameters $n, k, d_{min}$ along with $q$.

We can also state upper bounds on the \emph{network protection codes}~\cite{macwilliams77,huffman03}. The Singleton bound on the \emph{network protection code} parameters are stated as follows. Let $t$ be the number of failure that the code can protect.
\begin{eqnarray}\label{eq:singletonbound} t\leq n-k\end{eqnarray}
The equality in this bound will hold if the size of the used finite field is
greater than $n-t$.

One can also state the  Hamming bound in the network code parameters as follows.
%
\begin{eqnarray}\label{eq:hamming}
\sum_{i=0}^{\lfloor (d_{min}-1)/2\rfloor} \binom{n}{i} \Big (q-1\Big)^i\leq q^{n-k}
\end{eqnarray}

For the binary Hamming bound of $m$ protection paths

\begin{eqnarray}
\sum_{i=0}^{\lfloor t/2\rfloor} \binom{n}{i} \leq 2^{m}
\end{eqnarray}

We have the following lemma on the minimum number of protection paths of \emph{network protection code} parameters.
\begin{lemma}
\begin{eqnarray}
m \geq \max \Big\{d_{min}-1, \log_q \Big(\sum_{i=0}^{\lfloor t/2\rfloor} \binom{n}{i} \Big (q-1\Big)^i\Big) \Big\}
\end{eqnarray}
\end{lemma}
\begin{proof}
The proof is a direct consequence from the Singleton and Hamming bounds. Applying Equations~(\ref{eq:singletonbound}) and (\ref{eq:hamming}) gives the result.
\end{proof}
\section{Tables of Best Known Protection Codes}\label{sec:bestcodes}

\begin{table}[t]
\caption{Best known  \emph{network protection  codes} against up to four link failures. Such codes can be punctured, extended, or shortened to obtain the required length as shown in Section~\ref{sec:bestcodes}.}
\label{table:bchtable2}
\begin{center}
\begin{tabular}{|l|l|l|l|}
\hline    n& m &code&type  \\
 \hline
 &&&\\
15&7&$[15, 8, 5]_2$& Hamming code\\
19&8&$[19,11,5]_2$&Lengthening  Hamming-Preparata code\\
20&11&$[20,9,5]_2$&Lengthening Hamming-Preparata code \\
23&9&$[23, 14, 5]_2$& Linear code\\
31&10&$[31, 21, 5]_2$&BCH code \\
33&10&$[33, 23, 5]_2$&Linear code \\
35&13&$[35, 22, 5]_2$& Shorting Preparata code\\
63&11&$[63,52,5]_2$&Preparata code\\
70&12&$[70,58,5]_2$&Lengthening Hamming-Preparata code\\
81&13&$[81,68,5]_2$&Linear code\\
128&14&$[128, 114,5]_2$&BCH code\\
135&18&$[135,117,5]_2$&Shorting Preparata code\\
\hline
\end{tabular}
\end{center}
\end{table}
In this section we investigate which codes are suitable for network self-healing against link failures. We will present several \emph{network protection codes} with given generator matrices and exact parameters. The proposed codes are not necessarily optimal, i.e. they do not saturate the Singleton bound.
The classical Singleton bound is given by
\begin{eqnarray}\label{eq:Singleton}
k\leq n-d_{min}+1
\end{eqnarray}
This bound shows that the number of protection paths must be at least $d_{min}-1$, i.e., $m \geq d_{min}-1$. The equality of this inequality occurs in case of a single path failure.

We notice that all senders do not participate in the encoding vectors. This means that the proposed codes are suitable for the general protection case where a set of working paths is protected by a protection path. This will reduce our proposed codes to be also used for network protection using p-cycle~\cite{kamal07b,kamal06a}.

The codes shown in Table~(\ref{table:singledoublefailures}) are used to protect against single and double link failures using their symmetric generator matrices. Also, the codes in Table~~(\ref{table:bchtable2}). Table~(\ref{table:bchtable3}) presents the best known BCH codes for arbitrary minimum distance  over $\F_2$.
\begin{table}[t]
\caption{Families of  BCH codes that can be used as  \emph{network protection codes} against  link failures.}
\label{table:bchtable3}
\begin{center}
\begin{tabular}{|c|c|c|}
\hline   n & m& BCH Code  \\
 \hline
 &&\\
 15&4&$[15,11,3]$\\
 15&7&$[15,8,4]$\\
 15&8&$[15,7,5]$\\
 31&5&$[31,26,3]$\\
 31&5&$[31,26,3]$\\
 31&10&$[31, 21, 5]$\\
 31&15&$[31,16,7]$\\
 31&10&$[31,11,11]$\\
 31&25&$[31,6,15]$\\
127&14&$[127,113,5]$\\
127&49&$[127,78,15]$\\
127&21&$[127,106,7]$\\
127&50&$[127,77,27]$\\
  \hline
\end{tabular}
\end{center}
\end{table}

Given a NPC with parameters $[n,k,d_{min}]$, one can possibly obtain a new NPC
by shortening, extending, or puncturing this code.
If there is an NPC $\C$ with parameters $[n,k,d_{min}]_2$, then by
\begin{inparaenum}[i)]
\item shortening $\C$ yields a code with parameters $[n-1,k-1,d_{min}]2,$
\item puncturing $\C$ yields a code with parameters $[n-1,k,d_{min}-1]_2,$
\item    appending $\C$ yields a code with parameters $[n+1,k,d_{min}+1]_2,$
\item   extending $\C$ yields a code with parameters $[n+1,k+1,d_{min}]_2.$
\end{inparaenum}

For example, if there is a BCH Hamming code with parameters $[15,11,3]]_2$,  then there must be  codes with parameters $[14,10,3]_2$ (by shortening), $[14,11,2]_2$ (by puncturing), $[16,11,4]_2$ (by appending), $[16,12,3]_2$ (by extending). The interested readers might consult textbooks in classical coding theory for further  propagation rules~\cite{huffman03,macwilliams77}.
\subsection{Illustrative Examples}
\begin{example}
Consider a BCH code $C$ with parameters $[15,11,3]_2$  that has
designed distance $3$ and generator matrix $G$ given by:

\begin{eqnarray} \left[\begin{array}{p{0.1cm}p{0.1cm}p{0.1cm}cc ccccc cccccc}
1& 0& 0& 0& 0& 0& 0& 0& 0& 0& 0& 1& 1& 0& 0\\
0& 1& 0& 0& 0& 0& 0& 0& 0& 0& 0& 0& 1& 1& 0\\
0& 0& 1& 0& 0& 0& 0 &0 &0& 0& 0& 0& 0& 1& 1\\0& 0& 0& 1& 0 &0 &0& 0&
0& 0& 0& 1& 1& 0& 1\\0& 0& 0& 0& 1& 0& 0& 0& 0& 0& 0& 1& 0& 1& 0\\0
&0& 0& 0& 0& 1& 0& 0& 0& 0& 0& 0& 1& 0& 1\\0& 0& 0& 0& 0& 0& 1& 0&
0& 0& 0& 1& 1& 1& 0\\0& 0& 0& 0& 0& 0& 0& 1& 0& 0& 0& 0& 1& 1& 1\\0&
0& 0 &0& 0& 0& 0& 0& 1& 0& 0& 1& 1& 1& 1\\0& 0& 0& 0& 0& 0& 0& 0& 0&
1& 0& 1& 0& 1& 1\\0& 0& 0& 0& 0& 0& 0& 0& 0& 0&1 &1 &0 &0& 1
\end{array}\right]
\end{eqnarray}

The code $C$ over $\F_2$ can be used to recover from two link failures since its minimum distance is $3$. One can puncture, shorten, or extend this code to obtain the required code length, which determines the total number of disjoint connections.
In this example we have $15$ connections, and $11$ primary working paths. Furthermore, the
links $L_{12},L_{13},L_{14},L_{15}$ will carry encoded data.
The matrix $G$  presents the construction of NPC, and the senders that will send encoded and plain data.
\end{example}

\begin{example}
The code $C$ has parameters $[15, 8, 4]_2$ and generator
matrix $G$ given by:
\begin{eqnarray} \left[\begin{array}{p{0.1cm}p{0.1cm}p{0.1cm}cc ccccc cccccc}
1 &0 &0& 0& 0& 0& 0& 0& 1& 1& 0& 1& 0& 0& 0\\0& 1 &0 &0 &0 &0 &0 &0&
0& 1& 1& 0& 1& 0& 0\\0& 0& 1& 0& 0& 0& 0& 0& 0& 0& 1& 1& 0& 1& 0\\0&
0& 0& 1& 0& 0& 0& 0& 0& 0& 0& 1& 1& 0& 1\\0& 0& 0& 0& 1& 0& 0& 0& 1&
1& 0& 1& 1& 1& 0\\0& 0& 0& 0& 0& 1& 0& 0& 0& 1& 1& 0& 1& 1& 1\\0& 0&
0& 0& 0& 0& 1& 0& 1& 1&1 &0 &0& 1& 1\\0& 0& 0& 0& 0& 0& 0& 1& 1& 0&
1& 0& 0& 0& 1
\end{array}\right]
\end{eqnarray}
This means that all senders $s_1,\ldots,s_8$ will send plain data over
the working connection $L_1,\ldots,L_8$. Also, the senders
$s_9,\ldots,s_{15}$ will send encoded data over the protection paths
$L_{9},\ldots,L_{15}$.  In this encoding scheme, the connection
$L_9$ will carry encoded data from $s_1,s_5,s_7$ and $s_8$.
\end{example}

\section{ILP Formulation}
\label{sec:ilp}

The problem of finding link disjoint paths between pairs of nodes in a
graph is known to be an NP-complete problem \cite{Vygen95}.
Hence, even finding the working paths in this problem is hard.
We therefore introduce an Integer Linear Program (ILP) for solving the
reduced capacity network coding-based protection
problem introduced in this paper.

The purpose of the ILP is to find a feasible provisioning for groups
of connections, such that:
\begin{itemize}
\item
The paths used by a group of connections protected together are mutually
link disjoint.
\item
There is a circuit, {\bf S}, which connects the sources of all connections
protected together, and this circuit is link disjoint from the working
paths.
The {\bf S} circuit is used to exchange source data units in order to
form the linear combination of data units to be sent on the path used
for that purpose.
\item
There is a circuit, {\bf R}, which connects the receivers of all connections
protected together, and this circuit is link disjoint from the working
paths.
The {\bf R} circuit is used by the receivers to recover from lost data units
due to a failure.
\item
The total number of links used by the working paths, the {\bf S} circuit and
the {\bf R} circuit is minimal.
\end{itemize}

We assume that the number of channels per
span is not upper bounded, i.e., the
network is uncapacitated.

The following table defines the input parameters to the ILP:\\
\begin{tabular}{l p{2.8in}}
$N$ & number of connections\\
$s_h$ & source of connection $h$\\
$r_h$ & destination of connection $h$\\
$\delta^{hl}$ & a binary indicator which is equal to 1 if connections
  $h$ and $l$ have the same destination\\
$\gamma^{hl}$ & a binary indicator which is equal to 1 if connections
  $h$ and $l$ have the same source\\
\end{tabular}

The variables used in the formulation are given below:\\
\begin{tabular}{l p{2.8in}}
$n^{hl}$ & binary variable which is 1 if and only if connections $h$
  and $l$ are protected together\\
$z^h_{ij}$ & binary variable which is 1 if and only if connection $h$ uses
link ($i,j$) on the working path\\
$p^h_{ij}$ & binary variable which is 1 if and only if connection $h$ uses
link ($i,j$) on its {\bf S} circuit\\
$q^h_{ij}$ & binary variable which is 1 if and only if connection $h$ uses
link ($i,j$) on its {\bf R} circuit\\
$b^h_{i,j}$ & binary variable which is 1 if connection $h$ uses
link ($i,j$) on its backup path\\
\end{tabular}
\begin{tabular}{l p{2.8in}}
$P^{hl}_j$ & binary variable, which is 1 if and only if the {\bf S}
circuits for connections $h$ and $l$ share a node, $j$
  (required if $n^{hl} =1$).\\
$Q^{hl}_j$ & binary variable, which is 1 if and only if the {\bf R}
circuits for connections $h$ and $l$ share a node, $j$
  (required if $n^{hl} =1$ and $\delta^{hl}=0$).\\
$P^h$ & binary variable, which is 1 if and only if connection $h$ is
protected with another connection that has a source different
than that of $h$  (this variable is important since if $h$ is
not protected with another such connection, there is no need for the {\bf S}
circuit).\\
$Q^h$ & binary variable, which is 1 if and only if connection $h$ is
protected with another connection that has a destination different
than that of $h$  (this variable is also important since if $h$ is
not protected with another such connection, there is no need for
the {\bf R} circuit).\\
${\cal P}^{hl}_{ij}$ & binary variable which is 1 if and only if
connections $h$ and $l$ are protected together, and share link ($i,j$)
on the {\bf S} circuit.\\
${\cal Q}^{hl}_{ij}$ & binary variable which is 1 if and only if
connections $h$ and $l$ are protected together, and share link ($i,j$)
on the {\bf R} circuit.\\
\end{tabular}
\begin{tabular}{l p{2.8in}}
$\pi^h_{i,j}$ & binary variable which is equal to 1 if connection $h$ is the
lowest numbered connection, among a number of jointly protected connections,
to use link $(i,j)$ on its {\bf S} circuit (used in computing the cost of the {\bf S} circuit).\\
$\theta^h_{i,j}$ & binary variable which is equal to 1 if connection $h$ is the
lowest numbered connection, among a number of jointly protected connections,
to use link $(i,j)$ on its {\bf R} circuit (used in computing the cost of the {\bf R} circuit).\\
$\beta^h_{i,j}$ & binary variable which is equal to 1 if the secondary
protection path for connection $j$ uses link $(i,j)$.\\
\end{tabular}

\textbf{Minimize:}

\[
\sum_{i,j,h}  (z^h_{i,j} + \beta^h_{i,j} + 0.5 \pi^h_{i,j}  + 0.5 \theta^h_{i,j} )
\]
In the above, the summation is the
cost of the links used by the connections'
working paths and the {\bf S} and {\bf R} circuits.
It also includes the cost of a secondary circuit for 1+1 protection,
in case network coding-based protection cannot be used.
The calculation of these cost factors will be explained using the
constraints below.

\textbf{Subject to:}\\
The following constraints are enforced in the working and protection paths.

\bigskip

\noindent \textbf{I- Constraints on working paths:}
\begin{eqnarray}
&&z^h_{i,s_h} = 0 ~~~ \forall h,~ i \neq s_h \label{eqn:source1-working} \\
&&z^h_{r_h,j} = 0 ~~~ \forall h,~j \neq r_h \label{eqn:destination1-working} \\
&& \sum_{i \neq s_h} z^h_{s_h,i} = 1  ~~~ \forall h \label{eqn:source2-working} \\
&& \sum_{i \neq r_h} z^h_{i,r_h} = 1  ~~~ \forall h \label{eqn:destination2-working} \\
&& \sum_i z^h_{ij} = \sum_i z^h_{ji}  ~~~ \forall h, ~j \neq s_h,
  ~r_h \label{eqn:flow-working} \\
&& z^h_{ij} + z^h_{ji} + z^l_{ij} + z^l_{ji} + n^{hl} \leq 2 ~~~
  \forall h, l, i, j
\label{eqn:joint-protection}
\end{eqnarray}

Equations (\ref{eqn:source1-working}), (\ref{eqn:source2-working}),
(\ref{eqn:destination1-working}) and  (\ref{eqn:destination2-working})
ensure that the traffic on the working path
is generated and consumed by the source and destination nodes, respectively.
Equation (\ref{eqn:flow-working}) guarantees flow continuity on the working
path.
Equation (\ref{eqn:joint-protection}) ensures that the working paths of two connections which are protected together are link disjoint.
Since a working path cannot use two links in opposite directions on the
same span (or edge in the graph), then two connections which are
protected together cannot use the same span either in the same, or
opposite directions.
Such a condition is included in Equation
(\ref{eqn:joint-protection}).

\bigskip

\noindent \textbf{II- Constraints on secondary protection circuits:}
\begin{eqnarray}
&&b^h_{i,s_h} = 0 ~~~ \forall h,~ i \neq s_h \label{eqn:source1-backup} \\
&&b^h_{r_h,j} = 0 ~~~ \forall h,~j \neq r_h \label{eqn:destination1-backup} \\
&&\sum_{i \neq s_h} b^h_{s_h,i} = 1  ~~~ \forall h \label{eqn:source2-backup} \\
&&\sum_{i \neq r_h} b^h_{i,r_h} = 1  ~~~ \forall h \label{eqn:destination2-backup} \\
&&\sum_i b^h_{ij} = \sum_i b^h_{ji}  ~~~ \forall h, ~j \neq s_h,
  ~r_h \label{eqn:flow-backup} \\
&&\beta^h_{ij} \geq b^h_{ij} - \sum_l n^{hl} ~~~ \forall h, ~l,~i,~j
  \label{eqn:flow-secondary} \\
&&\beta^h_{ij} + z^h_{ij} \leq 1 ~~~ \forall h,~i,~j
\label{eqn:working-secondary-disjoint}
\end{eqnarray}
The above constraints evaluate the cost of the secondary protection paths used
for 1+1 protection.
There are two sets of variables in the calculation of this cost.
The first one is the $b^h_{ij}$ variables, which are evaluated in Equations
(\ref{eqn:source1-backup})-(\ref{eqn:flow-backup}) using exactly the same
way the $z^h_{ij}$ variables are evaluated.
However, the cost that goes into the objective function depends on whether connection
$h$ is protected with another connection using network coding or not.
The variables which evaluate this cost are the $\beta^h_{ij}$ variables, and are
evaluated in Equation (\ref{eqn:flow-secondary}), which makes it equal to $b^h_{ij}$
only if the connection is not protected with another connection.
Finally, Equation (\ref{eqn:working-secondary-disjoint}) makes sure that the working
and the used secondary paths are link disjoint.

\bigskip
\noindent
\textbf{III- Constraints on {\bf P} circuits:}
\begin{eqnarray}
&&P^h \geq n^{hl} -\gamma^{hl} ~~~ \forall h, l \label{eqn:Ph} \\
&&\sum_{i} p^h_{s_hi} = P^h ~~~ \forall h,~ i \neq s_h
\label{eqn:source1-primary-protection} \\
&&\sum_i p^h_{is_h} = P^h ~~~ \forall h,~i \neq s_h
\label{eqn:destination1-primary-protection} \\
&& \sum_i p^h_{ij} = \sum_i p^h_{ji}  ~~~ \forall h, ~j \label{eqn:flow-primary} \\
&& z^h_{ij} + \frac{p^h_{ij} + p^h_{ji}}{2} \leq 1 \label{eqn:working-primary-disjoint} ~~~ \forall h, i, j\\
&& z^h_{ij} + \frac{p^l_{ij} + p^l_{ji}}{2} + n^{hl} \leq 2
\label{eqn:working-primary-disjoint-other} ~~~ \forall h, i, j\\
&& \sum_{i} ( p^h_{ij} + p^l_{ij} ) \geq 2 P^{hl}_j ~~~\forall h, l, j
\label{eqn:primary-joint-node-tail}\\
&& \sum_{i} ( p^h_{ji} + p^l_{ji} ) \geq 2 P^{hl}_j ~~~\forall h, l, j
\label{eqn:primary-joint-node-head}\\
&& \sum_{j} P^{hl}_j \geq n^{hl} -\gamma^{hl} ~~~\forall h,l
\label{eqn:primary-joint-node-enforced}
\end{eqnarray}

Equation (\ref{eqn:Ph}) ensures that the source of connection, $h$ will be
connected to a {\bf S} circuit only if it is jointly protected with
another connection, $l$. However, there is one exception to this case,
which is the case in which the two connections $h$ and $l$ have the same
source. In this case, the {\bf S} circuit is not needed, and this is why
$\gamma^{hl}$ is subtracted from the right hand side of the equation.
Notice that if $h$ is protected together with another connection that has
a different source, then Equation (\ref{eqn:Ph}) will then require that a
{\bf S} circuit be used. Equations (\ref{eqn:source1-primary-protection})
and (\ref{eqn:destination1-primary-protection}) will ensure that traffic
leaves and enters $s_h$ using the {\bf S} circuit, only if it is jointly
protected with another connection that has a different source, i.e., when
$P^h = 1$. Equation (\ref{eqn:flow-primary}) guarantees connection $h$'s
flow continuity on the {\bf S} circuit. Equation
(\ref{eqn:working-primary-disjoint}) makes sure that the working path and
its {\bf S} circuit are link disjoint, while Equation
(\ref{eqn:working-primary-disjoint-other}) makes sure that if two
connections $h$ and $l$ are jointly protected, then the {\bf S} circuit of
$l$ must also be disjoint from the working path of connection $h$. Notice
that both of Equations (\ref{eqn:working-primary-disjoint}) and
(\ref{eqn:working-primary-disjoint-other}) allow a {\bf S} circuit to use
two links in opposite directions on the same span, and this is why the sum
of the corresponding link usage variables is divided by $2$ in both
equations.
Equations (\ref{eqn:primary-joint-node-tail}),
(\ref{eqn:primary-joint-node-head}) and
(\ref{eqn:primary-joint-node-enforced}) make sure that if two connections,
$h$ and $l$, are protected together ($n^{hl} = 1$), then their {\bf S}
circuits must have at least one joint node ($P^{hl}_j=1$ for some $j$).
However, similar to Equation (\ref{eqn:Ph}), a {\bf S} circuit is not
needed if the two connections have the same source, hence the subtraction
of $\gamma^{hl}$ from the right hand side of Equation
(\ref{eqn:primary-joint-node-enforced}).

Notice that in
the ILP formulation, the constraints implement the {\bf S} circuit as a set of paths,
such that there is a path
from each source back to itself.
However, the requirement of at least one joint node between every pair
of such paths as enforced by constraint (\ref{eqn:primary-joint-node-enforced}) will make sure
that the {\bf S} circuit takes the form of a tree.

\bigskip
\noindent \textbf{IV- Constraints on {\bf R} circuits:}
\begin{eqnarray}
&&Q^h \geq n^{hl} -\delta^{hl} ~~~ \forall h, l \label{eqn:Qh} \\
&&\sum_{i} q^h_{r_hi} = Q^h ~~~ \forall h,~ i \neq r_h
\label{eqn:source1-secondary-protection} \\
&&\sum_i q^h_{ir_h} = Q^h ~~~ \forall h,~i \neq r_h
\label{eqn:destination1-secondary-protection} \\
&& \sum_i q^h_{ij} = \sum_i q^h_{ji}  ~~~ \forall h, ~j \label{eqn:flow-secondar2y} \\
&& z^h_{ij} + \frac{q^h_{ij} + q^h_{ji}}{2} \leq 1 \label{eqn:working-secondary-disjoint2} ~~~ \forall h, i, j\\
&& z^h_{ij} + \frac{q^l_{ij} + q^l_{ji}}{2} + n^{hl} \leq 2
\label{eqn:working-secondary-disjoint-other} ~~~ \forall h, i, j\\
&& \sum_{i} ( q^h_{ij} + q^l_{ij} ) \geq 2 Q^{hl}_j ~~~\forall h, l, j
\label{eqn:secondary-joint-node-tail}\\
&& \sum_{i} ( q^h_{ji} + q^l_{ji} ) \geq 2 Q^{hl}_j ~~~\forall h, l, j
\label{eqn:secondary-joint-node-head}\\
&& \sum_{j} Q^{hl}_j \geq n^{hl} -\delta^{hl} ~~~\forall h,l
\label{eqn:secondary-joint-node-enforced}
\end{eqnarray}

Equations (\ref{eqn:Qh})-(\ref{eqn:secondary-joint-node-enforced}) are
similar to Equations
(\ref{eqn:Ph})-(\ref{eqn:primary-joint-node-enforced}), but they apply to
the destinations and to the {\bf R} circuit. Therefore, the variables
$P^h$, $\gamma^{hl}$, $p^h_{ij}$ and $P^{hl}_j$ are replaced by $Q^h$,
$\delta^{hl}$, $q^h_{ij}$ and $Q^{hl}_j$, respectively.

\textit{Constraints on joint protection:}

\begin{eqnarray}
&& n^{hl} + n^{lm} - 1 \leq n^{hm} ~~~ \forall h, l, m \label{eqn:multiple-connection-protection}
\end{eqnarray}

Equation (\ref{eqn:multiple-connection-protection}) makes sure that if
connections $h$ and $l$ are protected together, and connections $l$ and
$m$ are also protected together, then connections $h$ and $m$ are
protected together.

\noindent \textbf{V- Constraints for cost evaluation:}
\begin{eqnarray}
&&{\cal P}^{hl}_{ij} \leq \frac{p^h_{ij} + p^l_{ij} +n^{hl}}{3}
  ~~\forall i, j, h, l \label{eqn:calPhlij} \\
&& {\cal Q}^{hl}_{ij} \leq \frac{q^h_{ij} + q^l_{ij} + n^{hl}}{3}
  ~~\forall i, j, h, l \label{eqn:calQhlij} \\
&&\pi^l_{ij} \geq p^l_{ij} - \sum_{h = 1}^{l-1} {\cal P}^{hl}_{ij}
  ~~~\forall l, i, j \label{eqn:cost-primary}\\
&&\theta^l_{ij} \geq q^l_{ij} - \sum_{h = 1}^{l-1} {\cal Q}^{hl}_{ij}
  ~~~\forall l, i, j \label{eqn:cost-secondary}
\end{eqnarray}

Equations (\ref{eqn:calPhlij}), (\ref{eqn:calQhlij}),
(\ref{eqn:cost-primary}) and (\ref{eqn:cost-secondary}) are used to
evaluate the cost of the {\bf S} and {\bf R} circuits, which are used in
the objective function. Equation (\ref{eqn:calPhlij}) will make sure that
${\cal P}^{hl}_{ij}$ cannot be 1 unless connections $h$ and $l$ are
protected together and share link $ij$ on the {\bf S} circuit.  Equation
(\ref{eqn:calQhlij}) will do the same thing for the {\bf R} circuit.  Note
that both ${\cal P}^{hl}_{ij}$ and ${\cal Q}^{hl}_{ij}$ should be as large
as possible since this will result in decreasing the cost of the {\bf S}
and {\bf R} circuits, as shown in Equations (\ref{eqn:cost-primary}) and
(\ref{eqn:cost-secondary}). In equation (\ref{eqn:cost-primary}),
$\pi^l_{ij}$ for connection $l$ will be equal to 1 only if it is not
protected on link $ij$ with another lower indexed connection, and will be
equal to 0 otherwise. That is, it is the lowest numbered connection among
a group of jointly protected connections that will contribute to the cost
of the links shared by the {\bf S} circuit. $\theta^l_{ij}$ which is
evaluated by Equation (\ref{eqn:cost-secondary}) will also follow a
similar rule, but for the {\bf R} circuit.

\section{ILP Evaluation and Cost Comparison}
\label{sec:cost}

In this section results from the ILP formulation
developed in the previous section to evaluate the cost of provisioning
circuits to provide self-healing in autonomic networks using the
proposed \emph{network protection codes}.
The ILP was solved using the Cplex linear programming solver
\cite{cplex}.
We also compare the cost of provisioning NPC to that of provisioning
1+1 protection.
The cost of 1+1 protection is evaluated using
Bhandari's algorithm \cite{Bhandari99}.

We ran the ILP for various network topologies. The network topologies
are generated randomly. First, we consider a bidirectional network with
6 nodes and 9 edges along with 4 and 5 connections. Second,  we consider
a network with 8 nodes and 12  edges along with 4 and 6 connections.
Finally, we consider a network with 10 nodes and 20 edges, while
provisioning 4 and 6 connections.

The results shown in Table~(\ref{table:results}) indicate that the cost of
provisioning self-healing using
NPC is always lower than that using 1+1 protection, and the saving in
the protection resources can reach up to 30\%.
strategy. For example, consider a network with 8 nodes, 12,
and 6 connections.
The total cost of using the 1+1 strategy is 26,
while the total cost of using NPC is 21.
The total saving in resources in this case is close to 20\%.
However, the saving in the protection resources only is more than 30\%.
The advantage of using NPC over 1+1 protection may even improve further
with the size of the network.
For example, for the case of the network with 10 nodes, 20 edges, and 4
connections, the total cost of 1+1 protection is 16, while the total
cost of NPC is 12, which means a total saving of 25\%.
The saving in the protection resources is also 40\% in this case.

\begin{table}[t]
\caption{Cost comparison between 1+1 and 1+N protection for networks
  with $|V|=6, ~|E|=9$; $|V|=8, ~|E|=12$; and
  $|V|=10,~ |E|=20$.}
  \label{table:results}
\centerline{\begin{tabular}{|c|c|c c c|c c c|}
\hline
\hline
$|V|, |E|$ & $N$ & \multicolumn{3}{|c|}{1+1} & \multicolumn{3}{|c|}{NPC}\\
\cline{3-8}
&  & Total & Working & Spare & Total & Working & Spare\\
\hline \hline
& 4  & 15 & 6&  9& 14 &6& 8\\
\cline{2-8}
6, 9&5  &17  &6&11 &12 & 6 &6\\
\hline \hline
&  4 &  19&9&10  &  16&8& 8\\
\cline{2-8}
8, 12 &  6 & 26 &10& 16 &  21&11& 10\\
\hline \hline
&  4 & 16 &6& 10 &  12&6& 6\\
\cline{2-8}
10,20&6&23&10&13&19&9&10\\
\hline \hline
\end{tabular}
}
\end{table}

\section{Conclusions}\label{sec:conclusion}
We studied a model for recovering from network link failures using
network coding. We defined the concept of
\emph{network protection codes} to protect against a single link
failure, and then extended this concept and the techniques
to protect against $t$ link failures
using network coding and reduced capacity. Such \emph{protection codes}
provide self-healing  in autonomic networks with a reduced control and
management plane complexity.
We showed that the encoding
and decoding processes are simple and can be done in a dynamic way.
We also developed an ILP formulation to optimally provision
communication sessions and the circuits needed to implement NPC.
This formulation was then used to assess the cost of implementing this
strategy, and to compare it to the cost of using 1+1 protection.
It was shown that the use of NPC for self-healing has an advantage over
1+1 protection, in terms of the cost of connection and backup circuit
provisioning.

\bigskip


\end{document}